\documentclass[osajnl,twocolumn,showpacs,footinbib,superscriptaddress,10pt]{revtex4-1} 
\usepackage{amsmath,amssymb,graphicx}
\usepackage{soul}
\begin{document}

\title{Optical forces from evanescent Bessel beams, multiple reflections and Kerker conditions  in magnetodielectric spheres and cylinders}

\author{Juan Miguel  Au\~{n}\'{o}n}\email{Corresponding author: jmaunon@icmm.csic.es}
\author{Manuel  Nieto-Vesperinas}
\affiliation{Instituto de Ciencia de Materiales de Madrid, Consejo Superior
de Investigaciones Cientificas, Campus de Cantoblanco, Madrid 28049,
Spain.}

\begin{abstract}

In this work we address, first,  the optical force on a magnetodielectric particle on a flat dielectric surface due to an evanescent Bessel beam and, second, the effects on the force of multiple scattering with the substrate.  For the first question we  find analytical solutions showing that due to the interference of the excited electric and magnetic particle dipoles, the vertical force unusually pushes the object out from the plane.  The incident wavelength rules whether the illumination constitutes, or not, an optical trap. 
As for the second problem, we make a 2D  study with a single evanescent plane wave, and we present  the  Kerker conditions, (so far established for spheres), for magnetodielectric  cylinders; showing that in $p$-polarization those are practically reproduced and are  associated to minima  of the horizontal and vertical forces. 
\end{abstract}


Published in J. Opt. Soc. Am. A {\bf 31} 1984 (2014)

\maketitle 
\section{Introduction}
In the area of near-field optics and nanophotonics  the role of evanescent waves is crucial.  In the most simple configuration, namely that of total internal reflection (TIR) beyond a critical incidence angle $\theta_c$, there have been several studies on the mechanical action of evanescent waves on micro and nanoparticles in order to create traps and manipulate them in several ways \cite{Kawata92OLevanescent,Almaas95JosaB,Arias2002modeling,Chaumet2002PRL,Zelenina07Laser,VolpePRL96}. This question has acquired a renewed interest after the discovery of tractor beams \cite{chen2011optical,novitsky2011single,sukhov2011negative,Qiu2012tractorMaterialInde} capable of pulling particles by negative radiation pressure (NRP) as a consequence of the interference of electric and magnetic dipoles when both are resonantly excited in such bodies \cite{nieto2010optical,GarciaEtxarri11Anistropic}. However, in spite of the interest of optical manipulation of particles with subwavelength resolution \cite{grier2003revolution}, to date no detailed theory nor controlled experiments exist on the possibilities of both trapping and creating tractor evanescent  waves either by TIR or by excitation of surface plasmon polaritons (SPP) at flat surfaces of noble metals \cite{nietoOL10evanescentforce}. Only some  detailed calculations  \cite{Arias2003attractive,yangZhao} report on this problem, although  the particles addressed have a low refractive index and thus no strong magnetodielectric resonances. 

To tackle this question, in this paper we carry out a theoretical and numerical study on the optical forces exerted on a dielectric particle with magnetic response on a flat dielectric surface under TIR conditions.  Such kind of objects have triggered much interest  due to their inherent optical magnetism \cite{PengPRL98, vynck2009alldielectric,nietoJOSA011,nieto2012nature,FilonovAPL12}. Propagating 
Bessel beams \cite{Durnin87prl,Durnin87JosaA} have manifested as suitable for creating NRP on these bodies in absence of surfaces \cite{chen2011optical,novitsky2011single,Qiu2012tractorMaterialInde}. Nevertheless, as far as we know their mechanical response to  evanescent Bessel fields  in TIR or SPP configurations have never been discussed. This problem goes beyond the study initiated in  \cite{nietoOL10evanescentforce} for non-resonant magnetic particles since it has two sides: on the one hand, the scattering of an incident evanesecent Bessel beam on a resonant magnetodielectric particle placed on a dielectric plane at which TIR occurs, and on which no multiple interactions take place. On the other hand, evanescent fields convey interfaces at which they are created; hence if these are close to a particle with a rather large scattering cross section, which may occur at peak excitation of its electric and / or magnetic resonant modes, such multiple reflections and scatterings will unavoidably be present, and we do not know detailed studies where their effect on the photonic forces have been addressed.

Therefore, we shall first present in Section II an analysis (to bereferred to as AR) of the forces upon a Si sphere in air with magnetodielectric response to illumination, exerted by  an evanescent Bessel beam transmitted from a dielectric plane by TIR. This will ignore multiple scattering with the plane. However, in contrast with previous studies on the action of evanescent waves on non-resonant objects, it will be shown in Section III how at resonant wavelengths, and depending on the signs and relative phases of polarizabilities, evanescent Bessel beams can exert repulsive vertical forces on magnetodielectric bodies. This latter  effect  will be shown  to be assisted by the plane which gives rise to  the above mentioned multiple reflections.

 Indeed the electric and magnetic dipole modes excited on the particle, (characterized by its first electric and magnetic Mie resonances, respectively), involve large scattering cross sections under resonant illumination and hence such multiple scattering  in many real situations. However, this analysis  (to be be denoted as MR) will be done with 2D numerical computations  on an infinite cylinder, rather than with a sphere. Also, a single evanescent plane wave, rather than a Bessel beam, will be assumed to illuminate the object. 

We reason that 2D results should translate to 3D configurations because of the small differences  existing between scattering from  magnetodielectric spheres and cylinders under $p$-polarization, (i.e. with the incident electric vector in the plane of incidence).  To this end we shall present details  never discussed so far  on electromagnetic scattering  in magnetodielectric cylinders, showing, like in spheres \cite{Kerker83Electromagnetic}, suppression of the scattered radiation in either the backward or forward directions. The consequences of these effects on the optical forces will also be discussed.

\section{Optical forces from evanescent Bessel beams}
 Bessel beams are non-diffracting solutions of the homogeneous Helmholtz equation  \cite{Durnin87prl,Durnin87JosaA}. In the scalar formulation, their wavefunction is  $U\left(r,\phi,z\right)=J_{0}\left(qr\right)e^{im\phi}e^{i\beta' z}$,
with $\beta'^{2}=k^{2}-q^{2}$;  $k=nk_{0}$, $k_{0}=2\pi/\lambda$ and $n$ is the refractive index of the medium. $q$ and $\beta'$ denote the transversal and  longitudinal components of the
wavevector, respectively. If $q^{2} > k^{2}$, $\beta'$ becomes imaginary and the wave amplitude decays along
the $z-$direction. Here we shall consider  the electric and magnetic vectors of evanescent  Bessel beams created  by TIR on a dielectric plane $z=0$. We employ cylindrical coordinates $\left(r,\phi,z\right)$,
so that the beam axis is directed along $OZ$ and $\beta'=i\beta=i\sqrt{q^{2}-k^{2}}$  \cite{Novitsky08evanescentbessel,Ruschin98JosaAevanescentBB}. Assuming harmonic time
dependence, $\exp\left(-i\omega t\right)$, these vectors are:
\begin{eqnarray}
\mathbf{E}\left(\mathbf{r}\right) & = & e^{im\phi-\beta z}\left(E_{r}\hat{\mathbf{e}}_{r}+E_{\phi}\hat{\mathbf{e}}_{\phi}+E_{z} \hat{\mathbf{e}}_{z}\right),\label{E}\\
\mathbf{H}\left(\mathbf{r}\right) & = & e^{im\phi-\beta z}\left(H_{r}\hat{\mathbf{e}}_{r}+H_{\phi}\hat{\mathbf{e}}_{\phi}+H_{z}\hat{\mathbf{e}}_{z}\right),\label{H}
\end{eqnarray}
where $E_i$ and $H_i$ ($i=r,\phi,z$) are 

\begin{eqnarray}
E_{r} & = & -\frac{\beta}{q}c_{2}J_{m}^{'}\left(qr\right)-\frac{km}{q^{2}r}c_{1}J_{m}\left(qr\right)\label{Er},\\
E_{\phi} & = & -\frac{ik}{q}c_{1}J_{m}^{'}\left(qr\right)-\frac{i\beta m}{q^{2}r}c_{2}J_{m}\left(qr\right),\label{Ephi:}\\
E_{z} & = & J_{m}\left(qr\right)c_{2},\label{Ez}
\end{eqnarray}
and
\begin{eqnarray}
H_{r} & = & -\frac{\beta}{q}c_{1}J_{m}^{'}\left(qr\right)+\frac{km}{q^{2}r}c_{2}J_{m}\left(qr\right),\label{Hr}\\
H_{\phi} & = & \frac{ik}{q}c_{2}J_{m}^{'}\left(qr\right)-\frac{i\beta m}{q^{2}r}c_{1}J_{m}\left(qr\right),\label{Hphi}\\
H_{z} & = & J_{m}\left(qr\right)c_{1}.\label{Hz}
\end{eqnarray}
In these equations $J_m\left(qr\right)$ is the Bessel function of order $m$,  ($m=0,1,2...$), $J_{m}^{'}\left(qr\right)=\text{d}J_m/\text{d}qr$, and
the arbitrary complex coefficients $c_{1}$ and $c_{2}$ describe the amplitude and the phase of TE ($s$) and TM ($p$) polarized waves, respectively. 

Now we address the electromagnetic force   from  such an evanescent Bessel beam on interaction with a magnetodielectric particle. For  dipolar particles, i.e. those whose scattering properties are fully characterized by their induced electric and magnetic dipoles \cite{GarciaEtxarri11Anistropic,nietoJOSA011,AunonPRA2013}, this force ${\bf F}$ is the sum: ${\bf F}=  {\bf F}^{e}+ {\bf F}^{m}+ {\bf F}^{em}$ on the excited electric dipole: ${\bf F}^{e}$, magnetic dipole: ${\bf F}^{m}$ and interference between them: ${\bf F}^{em}$. It can can be obtained substituting Eqs. (\ref{Er})-(\ref{Hz}) into Eqs. (42)-(44)
of reference \cite{nieto2010optical}. Assuming that the particle is in vacuum, the expressions for these electric, magnetic and interference forces along the radial coordinate are:
\begin{eqnarray}
&&F_{r}^{e}=\frac{1}{2}\text{Re}\left\{ \alpha_{e}^s\mathbf{E}\partial_{r}\mathbf{E}^{*}\right\}\nonumber \\
&&=\frac{1}{2}\text{Re}\lbrace
\alpha_{e}^s\rbrace e^{-2\beta z}\left(\frac{1}{8q}\left[\left(k^{2}\left|c_{1}\right|^{2}+\beta^{2}\left|c_{2}\right|^{2}\right)\right.\right.\nonumber\\
&&\times\left(\frac{2mJ_{m}}{qr}\left(J_{m-2}-J_{m+2}\right)+4J_{m}^{'}\left(J_{m-1}^{'}-J_{m+1}^{'}\right)\right)\nonumber\\
&&+k\beta\text{Re}\left\{ c_{1}c_{2}^{*}\right\} \left(\frac{4mJ_{m}}{qr}\left(J_{m-1}-J_{m+1}\right)\right.\nonumber\\
&&\left.\left.\left.+2J_{m}^{'}\left(J_{m-2}-J_{m+2}\right)\right)\right]+\left|c_{2}\right|^{2}qJ_{m}J_{m}^{'}\right),\label{Fer}
\end{eqnarray}
\begin{eqnarray}
&&F_{r}^{m}=\frac{1}{2}\text{Re}\left\{ \alpha_{m}^s\textbf{H}\partial_{r}\textbf{H}^{*}\right\}\nonumber \\
&&=\frac{1}{2}\text{Re}\lbrace\alpha_{m}^s\rbrace e^{-2\beta z}\left(\frac{1}{8q}\left[\left(k^{2}\left|c_{2}\right|^{2}+\beta^{2}\left|c_{1}\right|^{2}\right)\right.\right.\nonumber\\
&&\times\left(\frac{2mJ_{m}}{qr}\left(J_{m-2}-J_{m+2}\right)+4J_{m}^{'}\left(J_{m-1}^{'}-J_{m+1}^{'}\right)\right)\nonumber\\
&&-k\beta\text{Re}\left\{ c_{1}c_{2}^{*}\right\} \left(\frac{4mJ_{m}}{qr}\left(J_{m-1}-J_{m+1}\right)\right.\nonumber\\
&&\left.\left.\left.+2J_{m}^{'}\left(J_{m-2}-J_{m+2}\right)\right)\right]+\left|c_{1}\right|^{2}qJ_{m}J_{m}^{'}\right),\label{Fmr}
\end{eqnarray}
\begin{eqnarray}
&&F_{r}^{em}=-\frac{k^{4}}{3}\left[\text{Re}\lbrace \alpha_{e}^s\alpha_{m}^{s*}\rbrace\text{Re}\left(\mathbf{E}\times\mathbf{H}^{*}\right)_{r}\right.\nonumber\\
&&\left.-\text{Im}\lbrace\alpha_{e}^s\alpha_{m}^{s*}\rbrace\text{Im}\left(\mathbf{E}\times\mathbf{H}^{*}\right)_{r}\right]\nonumber\\
&&=-\frac{k^{4}}{3}\left[-\frac{2\beta m}{q^{2}r}J_{m}^{2}\text{Re}\lbrace\alpha_{e}^s\alpha_{m}^{s*}\rbrace\text{Im}\lbrace c_{1}c_{2}^{*}\rbrace\right.+\text{Im}\lbrace\alpha_{e}^s\alpha_{m}^{s*}\rbrace\nonumber\\
&&\times\left.\left(\frac{k}{q}J_{m}J_{m}^{'}\left(\left|c_{2}\right|^{2}-\left|c_{1}\right|^{2}\right)+\frac{\beta m}{q^{2}r}J_{m}^{2}\text{Re}\lbrace c_{1}c_{2}^{*}\rbrace\right)\right].\nonumber\\\label{Femr}
\end{eqnarray}
If the particle is dipolar in the wide sense \cite{GarciaEtxarri11Anistropic,nietoJOSA011}, its  magnetodielectric response, assumed it spherical, is characterized by its electric and magnetic polarizabilities
\begin{equation}
\alpha_{e}^{s}=i\frac{3}{2k^{3}}a_{1}^{s},\hspace*{1cm} \alpha_{m}^{s}=i\frac{3}{2k^{3}}b_{1}^s,
\label{polarizability_sphere}
\end{equation}
$a_{1}^s$ and $b_{1}^s$ standing for the electric and magnetic first Mie coefficients \cite{bohren1983absorption}. The superscript $s$ denotes that these quantities correspond to a sphere. For this type of illumination Eqs. (\ref{Fer})-(\ref{Fmr}) show that these radial components of the force are purely conservative (i.e. they are gradient forces) because the imaginary part of $\alpha_{e}^s$ or $\alpha_{m}^s$
does not contribute  \cite{nieto2010optical,aunon2012photonic}.  Nevertheless Eq. (\ref{Femr}) describes a non-conservative component associated to the electric-magnetic interference force  given by the product of these polarizabilities. 

Likewise we  obtain the force along the azimuthal $\hat{\mathbf{e}}_{\phi}$ direction: 
\begin{eqnarray}
F_{\phi} & = & F_{\phi}^{e}+F_{\phi}^{m}+F_{\phi}^{em}\nonumber \\
 & = & \frac{1}{2}\text{Im}\lbrace\alpha_{e}^s\rbrace\frac{m\left|\mathbf{E}\right|^{2}}{r}+\frac{1}{2}\text{Im}\lbrace\alpha_{m}^s\rbrace\frac{m\left|\mathbf{H}\right|^{2}}{r}\nonumber\\
 &&-\frac{k^{4}}{3}\text{Re}\lbrace\alpha_{e}^s\alpha_{m}^{s*}\rbrace\frac{km}{q^{2}r}J_{m}^{2}\left(qr\right)\left(\left|c_{1}\right|^{2}+\left|c_{2}\right|^{2}\right) \label{Fphi}.
\end{eqnarray}
Where
\begin{eqnarray}
&&\left|\mathbf{E}\right|^{2}/e^{-2\beta z}\nonumber\\
 &  &= \left(k^{2}\left|c_{1}\right|^{2}+\beta^{2}\left|c_{2}\right|^{2}\right)\left(\frac{J_{m}^{'2}\left(qr\right)}{q^{2}}+\frac{m^{2}J_{m}^{2}\left(qr\right)}{q^{4}r^{2}}\right)\nonumber\\
&& +\left|c_{2}\right|^{2}J_{m}^{2}\left(qr\right)+\frac{4k\beta m}{q^{3}r}J_{m}^{'}\left(qr\right)J_{m}\left(qr\right)\text{Re}\left\{ c_{1}c_{2}^{*}\right\}, 
\label{E2}
\end{eqnarray}
\begin{eqnarray}
&&\left|\mathbf{H}\right|^{2}/e^{-2\beta z}\nonumber\\
 & &= \left(k^{2}\left|c_{2}\right|^{2}+\beta^{2}\left|c_{1}\right|^{2}\right)\left(\frac{J_{m}^{'2}\left(qr\right)}{q^{2}}+\frac{m^{2}J_{m}^{2}\left(qr\right)}{q^{4}r^{2}}\right)\nonumber\\
 &&+\left|c_{1}\right|^{2}J_{m}^{2}\left(qr\right)-\frac{4k\beta m}{q^{3}r}J_{m}^{'}\left(qr\right)J_{m}\left(qr\right)\text{Re}\left\{ c_{1}c_{2}^{*}\right\}. 
\end{eqnarray}
The first two  terms of Eq. (\ref{Fphi}) are non-conservative 
and push the particle towards the positive azimuthal direction. The
third term due to the interference of the electric and magnetic
dipoles is either positive or negative depending on the sign of $\text{Re}\{\alpha_e^s\alpha_m^{s*}\}$. It is worth remarking that Eq. (\ref{Fphi}) is analogous to Eq. (2) of \cite{novitsky2011single} in the sense
that this azimuthal force will be negative if  the third term of  Eq. (\ref{Fphi}) dominates over
the first two. This is a new situation that deserves a discussion.

The most favorable situation to achieve such a negative azimuthal force is
when $\alpha_{e}^s=\alpha_{m}^s$ \cite{Qiu2012tractorMaterialInde}, this
corresponds to $a_{1}^s=b_{1}^s$, which pertains to the well-known {\it first Kerker's
condition} \cite{Kerker83Electromagnetic} to be be further discussed in Section III. In this case, and using the optical theorem for non-absorbing particles \cite{nietoJOSA011}, one has  that:  $\text{Im}\alpha_{e}^s=2k^{3}\left|\alpha_{e}^s\right|^{2}/3$. However, even including this latter case, it is easy to see from Eq.(\ref{polarizability_sphere}) that $F_{\phi}>0$ for any polarization, wavelength,
or  order $m$. Thus 
{\it the  azimuthal force from an  evanescent Bessel beam is always positive}; at least as long as there be no multiple scattering with the substrate interface.

On the other hand, the force components in the $z-$direction are:
\begin{equation}
F_{z}^{e}=\frac{-\beta}{2}\text{Re}\lbrace\alpha_{e}^s\rbrace\left|\mathbf{E}\right|^{2},\hspace{0.5cm} F_{z}^{m}=\frac{-\beta}{2}\text{Re}\lbrace\alpha_{m}^s\rbrace\left|\mathbf{H}\right|^{2},\label{Fez}
\end{equation}
\begin{eqnarray}
&&F_{z}^{em}=-\frac{k^{4}}{3}\left[\text{Re}\lbrace\alpha_{e}^s\alpha_{m}^{s*}\rbrace\text{Re}\left(\mathbf{E}\times\mathbf{H}^{*}\right)_{z}\right.\nonumber\\
&&\left.-\text{Im}\lbrace\alpha_{e}^s\alpha_{m}^{s*}\rbrace\text{Im}\left(\mathbf{E}\times\mathbf{H}^{*}\right)_{z}
\right]=\nonumber\\
&&-\frac{k^{4}}{3}e^{-2\beta z}\left[-\text{Re}\lbrace\alpha_{e}^s\alpha_{m}^{s*}\rbrace J_{m}J_{m}^{'}\frac{2m}{q^{3}r}\left(k^{2}-\beta^{2}\right)\text{Im}\left\{ c_{1}c_{2}^{*}\right\} \right.\nonumber\\
&&-\text{Im}\lbrace\alpha_{e}^s\alpha_{m}^{s*}\rbrace\left(\left(\frac{k\beta}{q^{2}}J_{m}^{'2}+\frac{k\beta m^{2}}{q^{4}r^{2}}J_{m}^{2}\right)\left(\left|c_{2}\right|^{2}-\left|c_{1}\right|^{2}\right)\right.\nonumber\\
&&\left.\left.+J_{m}J_{m}^{'}\frac{2m}{q^{3}r}\left(k^{2}-\beta^{2}\right)\text{Re}\left\{ c_{1}c_{2}^{*}\right\} \right)\right].\label{Femz}
\end{eqnarray}

Since the intensity $\left|\mathbf{E}\right|^{2}$
decays along the $z-$direction as $\exp\left(-2\beta z\right)$,
there is  a gradient force, one for the electric dipole and one for the magnetic, which will attract or repel the particle with respect to
the plane $z=0$ depending on whether the real part of the corresponding polarizability is positive or negative. Notice also from Eqs. (\ref{Femr}), (\ref{Fphi})
and (\ref{Femz}) that the non-conservative component $F_{z}^{em}$ due to the interference between both
dipoles is zero if and only if $\left|c_{2}\right|^{2}=\left|c_{1}\right|^{2}$
and $m=0$; (compare with the total azimuthal force $F_{\phi}$ which  is zero for $m=0$). 

In order to illustrate these results, let us consider a  magnetodielectric sphere of Si  illuminated in the near
infrared \cite{nietoJOSA011}, its refractive index being $n_{p}\simeq\sqrt{12}$.  In the range of size
parameter  $n_{p}kr_0<4.2$, ($r_0$ being the radius of the sphere),  it behaves as dipolar in the wide sense \cite{GarciaEtxarri11Anistropic}. 
\begin{figure}[t]
\begin{centering}
\includegraphics[width=\linewidth]{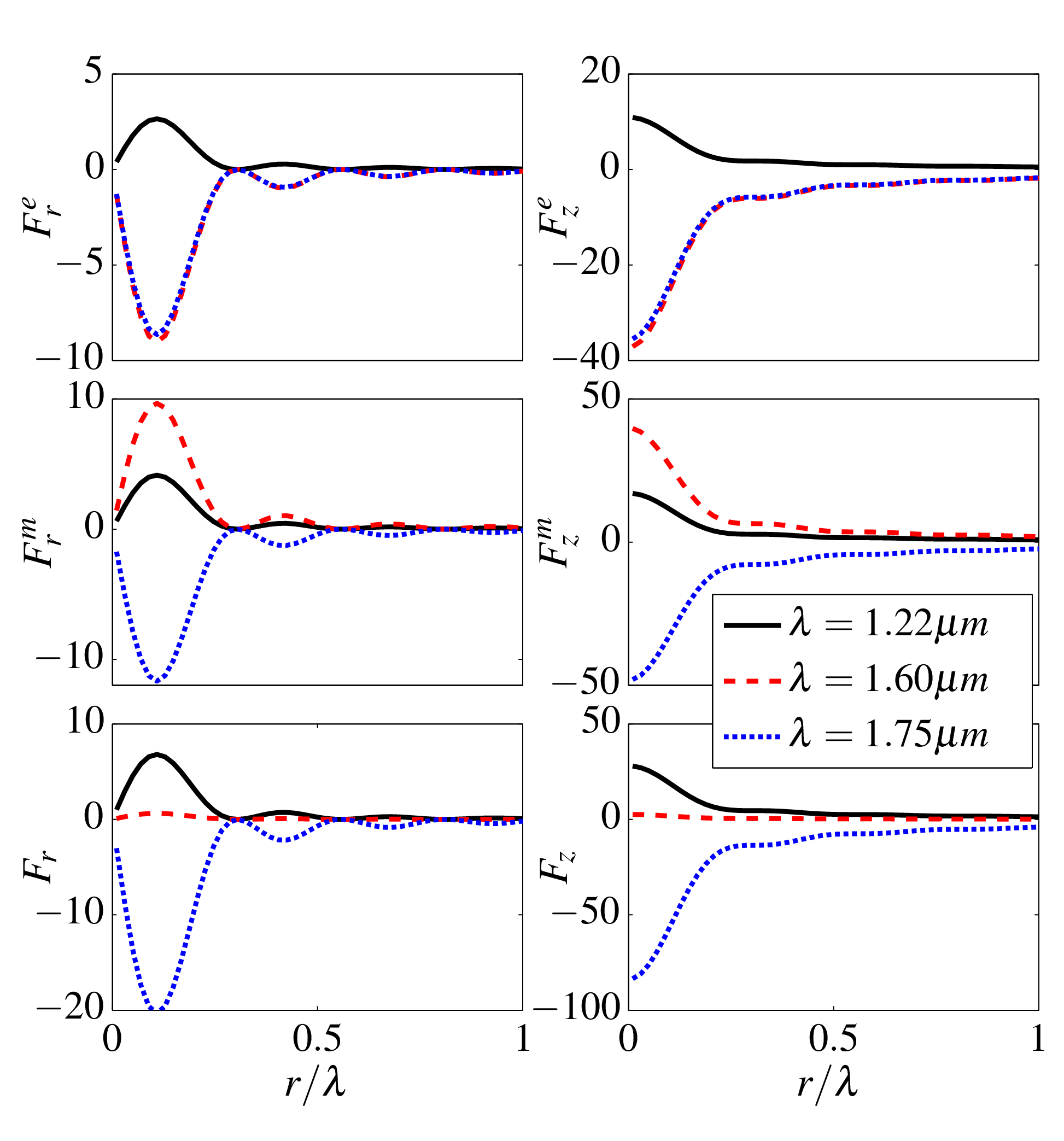}
\par\end{centering}
\caption{Optical force components, in arbitrary units, normalized to $\text{exp}(-2\beta z)$ on  a  Si sphere of $r_0=230nm$ from an evanescent Bessel beam with $m=0$. The horizontal rows show the electric and  magnetic forces and the sum of both respectively (the interference components $F_{r}^{em}$ and $F_{z}^{em}$ are zero).}  \label{fig:1}
\end{figure}

\begin{figure}[t]
\begin{centering}
\includegraphics[width=\linewidth]{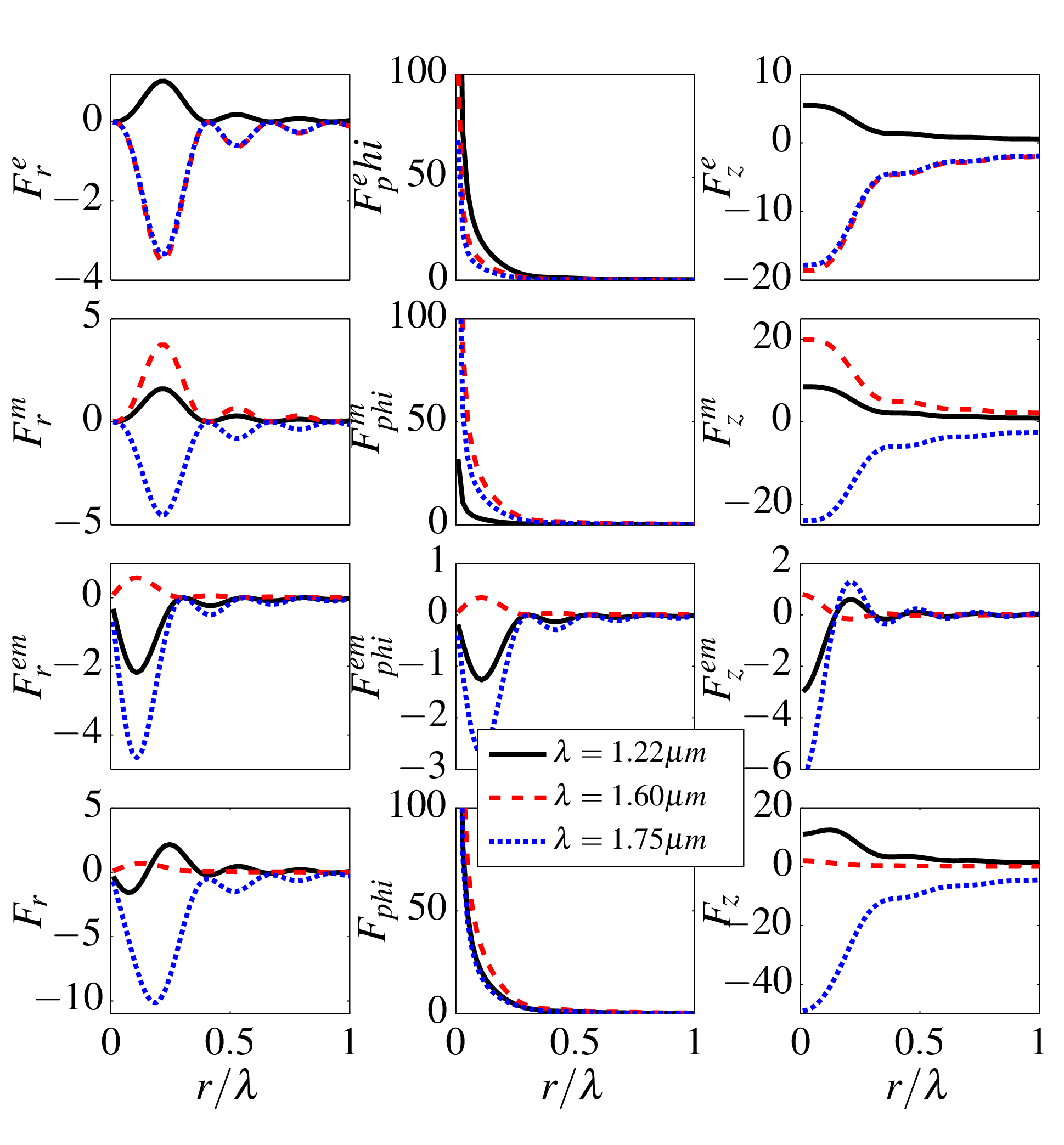}
\par\end{centering}
\caption{The same as Fig. \ref{fig:1} when  $m=1$. The horizontal rows show the electric, magnetic,  interference  and the total force for each component, respectively.} \label{fig:1_m1}
\end{figure}

\begin{figure}[t]
\begin{centering}
\includegraphics[width=\linewidth]{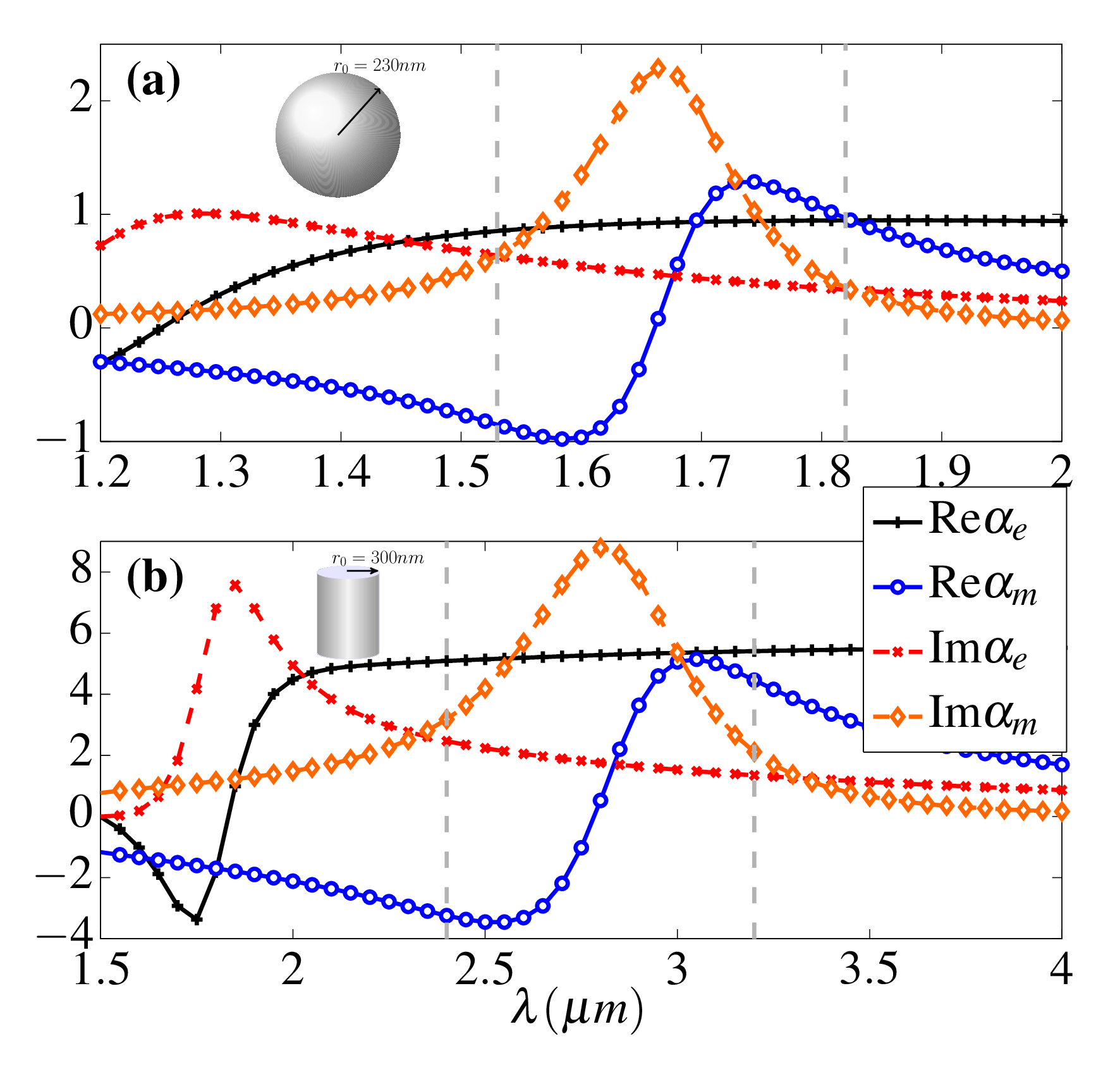}
\par\end{centering}
\caption{Real and imaginary parts of the electric ($\alpha_e$) and magnetic ($\alpha_m$) polarizabilities. (a) Sphere of radius $r_0=230nm$ (cf. Fig. 1(a) of \cite{nietoJOSA011}). (b) Cylinder of radius $r_0=300nm$. A plane polarized plane wave incides on the particles. The axis of the cylinder is along the magnetic vector. The results are normalized to $r_0^3$ for the sphere and to $r_0^2$ for the cylinder. The vertical broken lines indicate the 1st (right) and 2nd (left) Kerker condition wavelengths.} \label{polarizabilidad}
\end{figure}

Figs. \ref{fig:1} and \ref{fig:1_m1} illustrate the non-zero  components of the force from a  Bessel beam with $m=0$, ($c_1=1$, $c_2=i$, $q/k=2$), and $m=1$, ($c_{1}=1$, $c_{2}=i$, $q/k=2.0$), respectively,   for three  values of $\lambda$ and for a  Si sphere of $r_0=230nm$.  As shown above, all the interference and azimuthal forces are zero for $m=0$;
also in this case  Eqs. (\ref{Fer})-(\ref{Fmr}) and (\ref{Fez}), show that  the total force will be negative when  both $\text{Re}\{\alpha_{e}^s \}$ and $\text{Re}\{\alpha_{m}^s \}$ are positive, [cf. Fig. \ref{polarizabilidad} (a)]; then $F_r$ will pull the particle towards $r=0$, and $F_z$ will act as tractor force also pulling it towards the plane $z=0$. 

The case $m=1$ presents a richer behavior since now the non-conservative $F_{r}^{em}$ and $F_{z}^{em}$ components compete with those from the pure dipoles and can overcome them depending on the polarizabilites.  In this respect it is interesting to see in Fig. \ref{fig:1_m1} 
that while the purely electric and magnetic $r$ and $z$-components of the force from the  Bessel beam evanescent along $OZ$, are gradient and hence conservative forces, (and for $F_z$ this is in contrast with a  Bessel beam propagating in the $z$-direction which induces a scattering and thus no conservative force in this direction \cite{novitsky2011single}),  at certain values of the polarizabilities [cf.  Fig. \ref{polarizabilidad} (a)] $F_{r}^{em}$ and $F_{z}^{em}$  become attractive towards the center of the beam and the plane, respectively, thus  contributing to a resulting trapping force.  Therefore, \textit{tailoring the electric and magnetic polarizabilities}  [according  to Fig. \ref{polarizabilidad} (a)]  \textit{ the particle will, or will not, be trapped,  depending on the signs and relative phases of the polarizabilities at the illuminating wavelength }.  Trapping is clearly seen in  Figs. \ref{fig:1} and \ref{fig:1_m1} for $\lambda=1.75 \mu m$. Notice that actually the radial component of the force is similar to that from a propagating Bessel beam \cite{novitsky2011single} since the $r$ dependence is the same. 

Conversely, we observe in Fig. \ref{fig:1_m1} wavelengths where $F_{z}^{em}$ is opposite to the force from a pure dipole,  (we recall that the sign of ${\bf F}^{em}$ depends on the relative phases of $\alpha_{e}^{s}$ and $\alpha_{m}^{s}$),  and thus this interference component may contribute to changing the sign of the total vertical force. For example, at $\lambda=1.60 \mu m$ $F_{z}^{em}$  oposes to the attractive $F_{z}^{e}$, and together with $F_{z}^{m}$ produces a repulsive total vertical component $F_{z}$  even though the Bessel beam is evanescent along $OZ$. This is a consequence of the particle magnetodielectric resonant behavior and constitutes a novel situation opposite to those studied before \cite{Kawata92OLevanescent,Almaas95JosaB,Chaumet00electromagnetic} with evanescent plane waves that produce attractive vertical forces on passive non-resonant particles.




\section{Optical forces  in presence of  multiple reflections}
So far we have not addressed the interaction of the evanescent  beam with the surface that created it. This is an approach of frequent use in the literature. However, the existence of interfaces, inherent to such illuminating configurations, should introduce multiple scattering of the waves with the objects and the surfaces over which they are placed \cite{Arias2003attractive}. This should be even more prevalent under  particle resonance conditions here studied, since then its scattering cross-section is larger, thus giving rise to multiple reflections of the field with the interface. This  imposes a comparison of the ideal incident illumination model studied above, with that taking into account such multiple scattering. 

Without loss of generality, we now carry on numerical simulations in order to tackle this problem. The main goal is to check  the range of validity of the  non-multiple scattering  model for such large resonant particles as those with magnetodielectric response considered in this work. To make this task simpler, we consider only one plane wave component of the incident beam, which gives rise under TIR to one evanescent  component of the non-radiating  beam transmitted at $z=0$ and illuminating the particle.  This reasoning is warranted by the fact that the incident beam suffering TIR at $z=0$ and generating an evanescent beam,  is a superposition of plane waves, expressed by the angular spectrum representation of its wavefield. Since  the evanescent beam is characterized by a  a small angular dispersion of such non-radiating plane wave components \cite{nietolibro, mandel1995optical}, the deviations, or their absence, between the results  for a single plane wave component neglecting the field multiple reflections  and those  addressing them, would automatically be translated to the beam as it is built by superposing all such  plane waves.

Also, since full 3D computational simulations are numerically expensive, we  deal with the problem using 2D calculations, therefore we have chosen cylinders instead of spheres. In this connection notice that if one  addresses multiply reflected Bessel beams, a full 3D calculation with a sphere  should be done. 

\subsection{Scattering properties of magnetodielectric cylinders}
\label{Scattering properties of magnetodielectric cylinders}

Prior to entering in details of  multiple reflection effects, we discuss some scattering properties, never studied before, of a an infinitely long magnetodielectric dipolar cylinder illuminated by a plane polarized plane wave with incidence normal to the cylinder axis,  comparing them with those of a  sphere \cite{GarciaEtxarri11Anistropic, nietoJOSA011}. This is a case much  less addressed than that of a spherical particle.

The electric and magnetic polarizabilities of an infinite cylinder under $p$-polarization, i.e. with its axis along the ${\bf H}$-field and perpendicular to the incident wavector, are (notice now the superscript $c$): 
\begin{equation}
\alpha_{e}^{c}=i\frac{2}{\pi k^{2}}a_{1},\;\;\;\;\alpha_{m}^{c}=i\frac{1}{\pi k^{2}}a_{0}\label{alpha_cylinder}
\end{equation}
Details of these latter expressions can be found in the Appendix. $p$-polarization is the one under which the spectral behavior of the  polarizabilities of the cylinder are very similar to those of a sphere. A comparison between the polarizabilities of a Silicon sphere of radius $r_0=230nm$ and cylinder with $r_0=300nm$ is seen in Figs. \ref{polarizabilidad} (a) and (b), respectively.  [The results of Fig. 3(a) were obtained in \cite{nietoJOSA011} and are given here to facilitate comparison with those of Fig. 3(b)]. 

In addition, the angular distribution of scattered intensity for magnetodielectric spheres and  cylinders are \cite{bohren1983absorption,nieto2012nature}:
\begin{eqnarray}
I^{s}(\theta)&\propto & \left|a_{1}^{s}\cos\theta+b_{1}^{s}\right|^{2}, \nonumber\\
I^{c}(\theta)&\propto & \left|a_{0}^{c}+2a_{1}^{c}\cos\theta\right|^{2}, \label{Is}
\end{eqnarray}
respectively.  The  Mie coefficients $a_{n}^{s}$ ,$ b_{n}^{s}$, $a_{n}^{c}$
 can be found in e.g. \cite{bohren1983absorption}. Using Eqs. (\ref{polarizability_sphere}) and (\ref{alpha_cylinder}) the intensities can be written in terms of the polarizabilities and calculated from the values of Figs. \ref{polarizabilidad} (a) and (b).  

Two interesting scattering angles are those of the forward ($\theta=0$) and backward ($\theta=\pi$) directions. 

 In  forward scattering, the angular intensity is minimum  if  $\text{Re}{\alpha_e^{s,c}}=-\text{Re}{\alpha_m^{s,c}}$, $\text{Im}{\alpha_e^{s,c}}=\text{Im}{\alpha_m^{s,c}}$, (for spheres see \cite{Kerker83Electromagnetic, nietoJOSA011, nieto2012nature}).
 From Fig. \ref{polarizabilidad} one can see that the \textit{minimum forward intensity} is given at $\lambda\simeq1.53 \mu m$ (second Kerker condition) for the sphere and at $\lambda\simeq2.4 \mu m$ for the  cylinder. 
 
 
\begin{figure}[t]
\begin{centering}
\includegraphics[width=\linewidth]{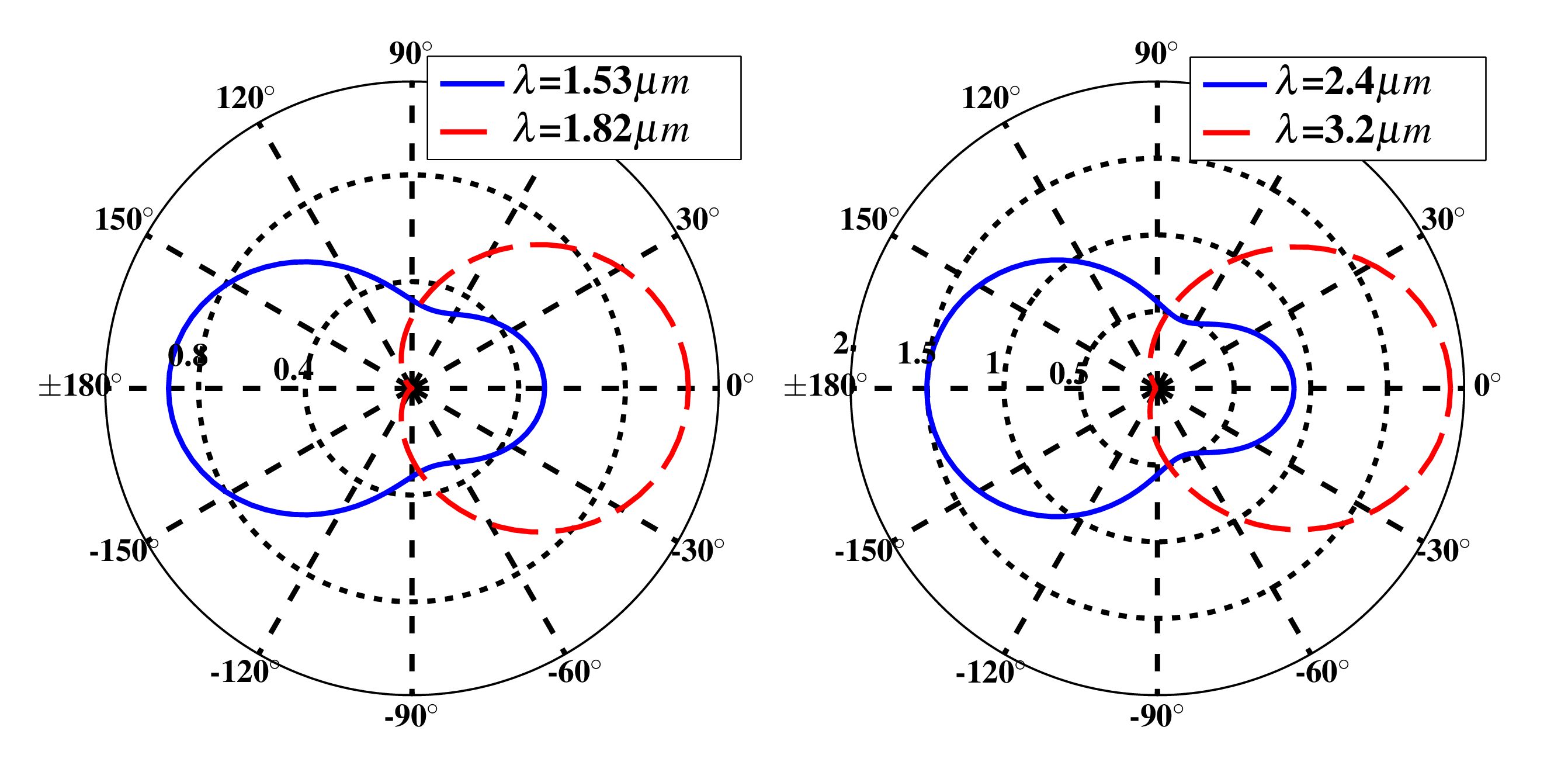}
\par\end{centering}
\caption{Angular distribution of intensity scattered from a magnetodielectric dipolar sphere (left) and cylinder under p-polarized illumination (right). The solid lines (blue) depict
the intensities at the wavelengths under which they become
zero (sphere) or near zero (cylinder) in $\theta=0$, whereas the
dashed (red, which appear multiplied by 1.52) lines plot the
intensities at the wavelengths at which they are near zero in
$\theta=\pi$  both for the sphere and cylinder.}
\label{angular_intensity}
\end{figure}

In the backward  direction, $\theta=\pi$, the intensity is zero if $a_{1}^{s}=b_{1}^{s}$ (i.e.  $\text{Re}\alpha_e^s=\text{Re}\alpha_m^s$, $\text{Im}\alpha_e^s=\text{Im}\alpha_m^s$) for  spheres, (first Kerker condition), \cite{Kerker83Electromagnetic, nietoJOSA011, nieto2012nature}, or if $a_{0}^{c}=2a_{1}^{c}$, ($\text{Re}\alpha_m^c=\text{Re}\alpha_e^c$, $\text{Im}\alpha_m^c=\text{Im}\alpha_e^c$) for   cylinders. For the sphere this latter condition is obtained at $\lambda\simeq1.82 \mu m$, being $I^{s}(\theta)$  exactly zero. However, due
to the factor 2 of the $a_{1}^{c}$ term in $I^{c}(\theta)$, [cf. the second Eq. (19)], such a zero backward intensity cannot be created by an infinitely long dipolar cylinder; nevertheless it can be minimum and close to zero. In our example this happens at $\lambda=3.2 \mu m$.

Fig. \ref{angular_intensity} depicts the angular scattered intensities for the above quoted wavelengths, demonstrating that these aforementioned anomalous scattering properties of magnetodielectric spheres at  $\theta = \pi$ and $\theta=0$ are almost  reproduced in magnetodielectric infinite cylinders under $p$-polarization.

\subsection{Effects of multiple reflections}
Based on the above discussed similarities of the scattering from dipolar spheres and cylinders under $p$-polarization,  we consider a propagating plane wave whose electric field  amplitude per sectional plane $y=0$ of the cylinder is $1V/m$, impinging  at an angle $\theta_0=34.2^o>\theta_c$ into the  plane interface $z=0$ that  separates  a dielectric of refractive index $n_1=2.58$  from air. 
We study its scattering with a magnetodielectric cylinder of radius $r_0$  placed in the air   at different distances $d$ from the plane, (i.e. with axis at $z=d+r_0$, $n_p=\sqrt{12}$ and $r_0=300nm$). 

In order to compare full numerical calculations of multiple reflection calculations  (MR) with  analytical results (AR) of an incident evanescent wave that ideally does not suffer multiple scattering with the $z=0$ plane, we first obtain the electromagnetic force ${\bf F}$ on the cylinder by means of Maxwell's stress tensor (MST) $\overleftrightarrow{\mathbf{T}}$, i.e., $\mathbf{F}=\int\overleftrightarrow{\mathbf{T}}\mathbf{n}dS$ ,  $\mathbf{n}$ being the unit outward normal to an arbitrary surface $S$ which encloses the particle \cite{jackson1998classical}. Notice that the fields from which the MST is obtained are total, i.e. incident plus  scattered: $\mathbf{E}_i+\mathbf{E}_s$ and $\mathbf{H}_i+\mathbf{H}_s$. 

The MR so obtained are exact within the resolution provided by the numerical procedure.   This is a finite element method (FEM), (RF module of Comsol 4.3). Details of how to calculate optical forces with the MST in a Comsol platform are to reported in \cite{Valdivia2012forces}.  Several grid samplings are done in order to check convergence of the computations. We then compare these MR calculations with the AR obtained from an   evanescent plane wave \cite{nietoOL10evanescentforce} that arises on transmission at $z=0$ of  the same $p$-polarized  plane wave propagating in the dielectric side. (We correct  misprints of Eqs. (2),  (5) and (6) of \cite{nietoOL10evanescentforce} for the  horizontal and vertical components of  ${\bf F}^{e}$ and ${\bf F}^{em}$.  The right side of (2) must be multiplied by -1. The factor $2\frac{K^2}{k^2}-1$ should be replaced by  $1$ in (5) and by $-1$ in (6). Notice that from those equations one has: $ F_x=   F_{x}^{e}+  F_{x}^{m}+ F_{x}^{em} \geq 0$.) 
\begin{figure}[t]
\begin{centering}
\includegraphics[width=\linewidth]{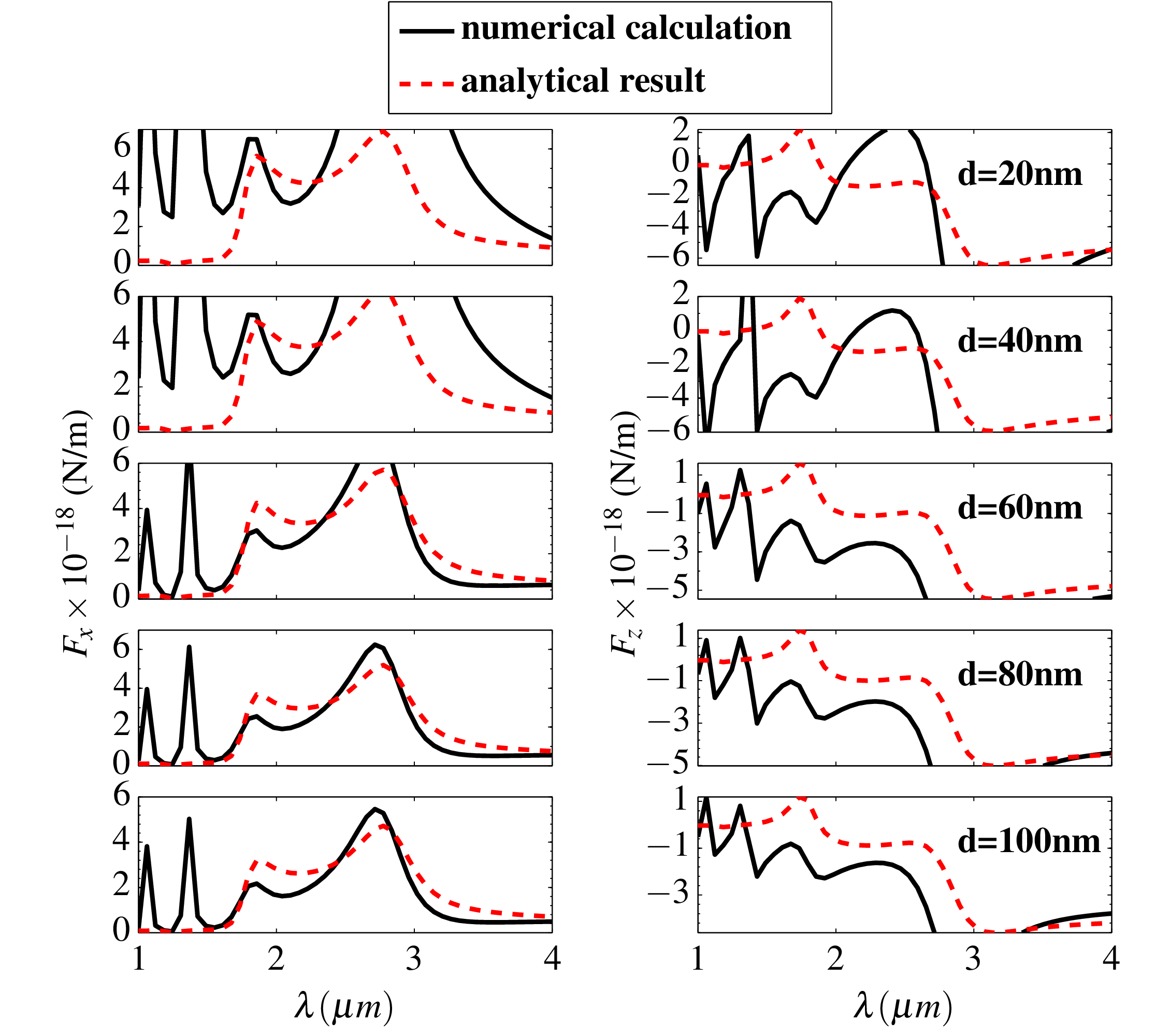}
\par\end{centering}
\caption{Optical forces, in $N/m$,  from a $p$-polarized evanescent plane wave created by TIR, on a Si cylinder ($r_0=300nm$) in air, with its axis at distance $d+r_0$ from a dielectric plane $z=0$, (see text). Left: lateral force $F_x$. Right: perpendicular force $F_z$. The solid  and dashed lines correspond to  MR and AR, respectively.   \label{Fig_r0_300}}
\end{figure}

Fig. \ref{Fig_r0_300}  shows the total force $\mathbf{F}(\textbf{r})=\mathbf{F}^e(\textbf{r})+\mathbf{F}^m(\textbf{r})+\mathbf{F}^{em}(\textbf{r})$ along the horizontal $x-$axis:  $F_x$,  and  its vertical component: $F_z$, exerted on the cylinder. $\lambda$ is the incident wavelength in air.  The two peaks of the numerical simulation MR for $\lambda<1.8\mu m$ represent multipolar modes of the cylinder and thus are not covered by the analytical calculation AR that assumes the particle scattering  being given by $a_0$ and $a_1$ only.  On the other hand, in the range $\lambda>1.8\mu m$, in which the scattering is due to the induced electric and magnetic dipoles,  peaked at $\lambda_1\simeq1.8\mu m$  and $\lambda_2\simeq2.7\mu m$, respectively, (on comparing the AR and MR dipolar peaks in Fig. 5 [see also Fig. 3(b)] notice their red-shift due to the presence of the plane), we observe a   much better agreement, even within the regions of  these resonance maxima, between the analytical (AR) and the full numerical (MR) calculations, providing that  $d/\lambda$ is not smaller than $r_0 /4$.

The departure between both models is evidently larger 
 as there is conversion by multiple scattering of a bigger proportion of evanescent waves (contributing to the gradient vertical force $F_z$) into radiative components (responsible of the radiation pressure in both the $x$ and $z$ directions) and viceversa \cite{Arias2003attractive,aunon2012photonic,AunonPRA2013}).  Of course as the incident wavelength grows, the dipolar response of the particle increases, the multiple reflection effect diminishes, both theoretical AR and numerical MR results tend to coincide with each other, and the gradient force  $F_z$ dominates and  remains attractive towards the plane $z=0$.

Notice that at any distance the radiation pressure $F_x$ is always positive, thus NRP cannot be obtained either with or without multiple reflections \cite{nietoOL10evanescentforce}. However it is remarkable that the presence of multiple scattering between the particle and the plane, obviously prevailing as $d$ decreases, turns the vertical force into pushing, thus loosing its gradient character. This latter change of the sign of $F_z$ is also yielded by the AR in the vicinity of the electric dipole resonance, even thought this is not confirmed by the MR. We also observe in Fig.  \ref{Fig_r0_300} that at the Kerker wavelengths, $F_x$ which follows the cylinder total scattering cross section  also mediated by the plane, presents minima, or near minima. This result for $F_x$ coincides with that for the radiation pressure exerted by an incident propagating plane wave \cite{nietoJOSA011}

As for $ F_z$  which follows both $Re \alpha_{e}$ and $Re \alpha_{m}$,  [see Fig. 3(b)], there is no minimum value in the region of the second Kerker wavelength although there is a minimum
in the neighboorhood of the first Kerker condition.

\begin{figure}[t]
\begin{centering}
\includegraphics[width=\linewidth]{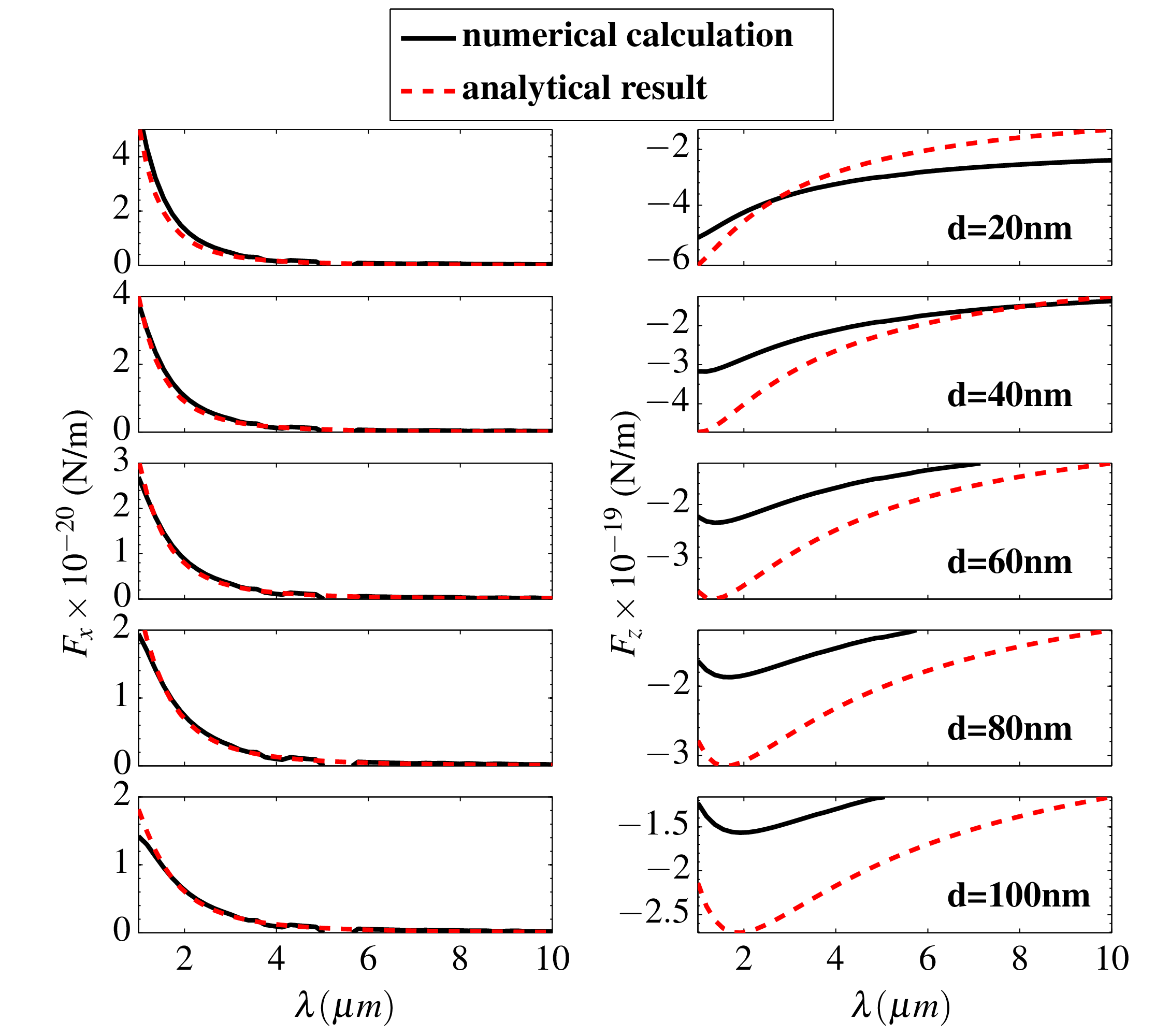}
\par\end{centering}
\caption{The same as in Fig. \ref{Fig_r0_300}  for a Si cylinder with $r_0=50nm$.Notice that the right column is one order of magnitude larger than the  left column since in general, for small $r_0/\lambda$, $\text{Re}\alpha_e>>\text{Im}\alpha_e$) \label{Fig_r0_50}}
\end{figure}

Multiple reflections decrease as the particle scattering cross section diminishes. A non-resonant smaller particle, (e.g. a  Si cylinder with $r_0=50nm$), suffers  negligible magnetic and electric-magnetic interference forces in the same range of $\lambda$ as before since then $\alpha_e\gg\alpha_m$. This is depicted in Fig. \ref{Fig_r0_50}. The left column exhibits a good agreement between the  analytical   and the numerical results.  Notice that now the non-conservative component $F_x$  is always positive, as expected, whereas $F_z$ behaves like that of a pure pulling gradient force from an evanescent wave \cite{Chaumet00electromagnetic, Arias2003attractive}. At this point it is worth remarking that the dependence on $\lambda$ of both $F_x$ and  $F_z$ is proportional to $\pm\text{exp}[4\pi z\sqrt{(n_1\sin\theta_0)^2-1)}/\lambda]/\lambda$, (the signs $+$ and $-$ apply for $F_x$  for $F_z$, respectively) \cite{Arias2003attractive,nietoOL10evanescentforce}, and this exponential has a maximum (minimum) at $\lambda=4\pi z \sqrt{(n_1\sin\theta_0)^2-1)}$. In particular, for e.g. $d=80nm$  the maximum (minimum) is at $\lambda=1.7\mu m$. This justifies the range of wavelengths there  addressed.

\section{Conclusions}

We have demonstrated that the vertical force exerted by an evanescent Bessel beam on a magnetic resonant particle presents unusual effects, similar, and reciprocal, to those of a propagating Bessel beam. While the latter may exert a NRP, the former contrary to evanescent waves studied so far, can produce a positive force on the body, pushing it outside the plane substrate. To this effect contributes the force arising from the interference of the electric and magnetic dipoles excited in the particle. The illumination wavelength also determines whether the system configurates, or not, a trap. In any case, the strong variation of the mechanical action with the frequency, also suggests its use for sorting purposes.

To deal in 2D  with the  effect of interactions of the scattered waves with the plane that creates the incident evanescent field, we have presented an analysis of the Kerker conditions, (previously established for spheres), for magnetodielectric infinitely long cylinders. Showing that under $p$-polarization these 2D objects practically reproduce the zero or near-zero backward, as well as the near-zero forward scattered intensities. 

These Kerker wavelengths are also close to those at which the magnitudes of the horizontal and vertical
forces present a minimum. This work also points out the importance of multiple scatterings and their consequences on the signs of the forces for these resonant particles.

\newpage
\appendix
\section*{Appendix: Electric and magnetic polarizabilities of infinitely long cylinders}

We  address the expressions for the electric
and magnetic polarizabilities for infinitely long magnetodielectric cylinders in terms
of the Mie coefficients. Firstly, from a general point of view, let
us express the electromagnetic field scattered (in SI units) by an electric ($\mathbf{p}$)
and magnetic (\textbf{$\mathbf{m}$}) dipoles placed at $\boldsymbol{\rho}_{0}$
 \cite{Yaghjian}
\begin{eqnarray}
\mathbf{E}_{s}\left(\mathbf{\boldsymbol{\rho}},\boldsymbol{\rho}_{0},\omega\right)&=&\mu_{0}\omega^{2}\overset{\leftrightarrow}{G}^{p}\left(\mathbf{\boldsymbol{\rho}},\boldsymbol{\rho}_{0},\omega\right)\mathbf{p}\left(\mathbf{\boldsymbol{\rho}}_{0},\omega\right) \nonumber\\
&-&\overset{\leftrightarrow}{G}^{m}\left(\mathbf{\boldsymbol{\rho}},\boldsymbol{\rho}_{0},\omega\right)\mathbf{m}\left(\mathbf{\boldsymbol{\rho}}_{0},\omega\right),\label{E_scatt_green}
\end{eqnarray}
where $\overset{\leftrightarrow}{G}^{p,m}\left(\mathbf{\boldsymbol{\rho}},\boldsymbol{\rho}_{0},\omega\right)$
are the 2D Green's functions of the system. In free-space propagation,
these Green's functions are: 
\begin{eqnarray}
\overset{\leftrightarrow}{G}^{p}\left(\mathbf{\boldsymbol{\rho}},\boldsymbol{\rho}_{0},\omega\right) & = & \left(\overset{\leftrightarrow}{I}+\frac{1}{k^{2}}\nabla\nabla\right)G_{0}\left(\mathbf{\boldsymbol{\rho}},\boldsymbol{\rho}_{0},\omega\right),\\
\overset{\leftrightarrow}{G}^{m}\left(\mathbf{\boldsymbol{\rho}},\boldsymbol{\rho}_{0},\omega\right) & = & \nabla G_{0}\left(\mathbf{\boldsymbol{\rho}},\boldsymbol{\rho}_{0},\omega\right)\times\overset{\leftrightarrow}{I,}
\end{eqnarray}
where $G_{0}\left(\mathbf{\boldsymbol{\rho}},\boldsymbol{\rho}_{0},\omega\right)$
is the 2D scalar Green's function 
\begin{equation}
G_{0}\left(\mathbf{\boldsymbol{\rho}},\boldsymbol{\rho}_{0},\omega\right)=\frac{i}{4}H_{0}\left(kr\right),
\end{equation}
with $r=\left|\mathbf{\mathbf{\boldsymbol{\rho}}-\boldsymbol{\rho}_{0}}\right|$ and $H_{0}\left(kr\right)=H_{0}^{(1)}\left(kr\right)$ denotes the first Hankel function. The main goal is to compare Eq. (\ref{E_scatt_green}) with the scattered electromagnetic field obtained from  the Mie scattering theory.

\subsection{S-polarization}

Let us consider an incident electromagnetic field impinging in an infinitely long cylinder of radius $r_0$ and refractive index $n_p=\sqrt{\varepsilon_p\mu_p}$ immersed in vacuum,  the electric vector being parallel to the cylinder axis, i.e., $\mathbf{E}_{in}=E_{0}e^{ikx}\hat{\mathbf{u}}_{z}$
$\left(k=\omega/c=2\pi/\lambda\right)$. For this polarization, the induced electric and
magnetic dipoles at the position of the particle are 
\begin{eqnarray}
\mathbf{p}\left(0,\omega\right) & = & \varepsilon_{0}\alpha_{e}^{s}E_{0}\hat{\mathbf{u}}_{z},\label{p}\\
\mathbf{m}\left(0,\omega\right) & = & \alpha_{m}^{s}H_{0}\hat{\mathbf{u}}_{y},\label{m}
\end{eqnarray}
where $\alpha_{e}^{s}$, $\alpha_{m}^{s}$ read \cite{LAKHTAKIA,madrazo,antonio,ChaumetPRB2000}
\begin{equation}
\alpha_{e}^{s}=\frac{\alpha_{e,0}^{s}}{1-\frac{ik^{2}\alpha_{0}^{s}}{4}}
\end{equation}
\begin{equation}
\alpha_{m}^{s}=\frac{\alpha_{m,0}^{s}}{1-\frac{i\alpha_{m,0}^{s}k^{2}}{8}},
\end{equation}
In the Rayleigh limit $k n_p r_0 <<1$, the static polarizabilities are $\alpha_{e,0}^{s}=\pi r_0^{2}\left(\varepsilon_{p}-1\right),\label{alpha_e_dynamic}$ $\alpha_{m,0}^{s}=2\pi r_0^{2}\frac{\mu_{p}-1}{\mu_{p}+1}.\label{alpha_m_dynamic}$
The superscript $s$ is understood as $s-$polarization. 

On the other hand, from  Mie theory the scattered field
from a infinite right circular cylinder is \cite{bohren1983absorption}
\begin{eqnarray}
&&\mathbf{E}_{s}\left(\mathbf{\boldsymbol{\rho}},\boldsymbol{\rho}_{0},\omega\right)=\nonumber\\
&&-E_{0}\sqrt{\frac{2}{\pi kr}}e^{-i\pi/4}e^{ikr}\left[b_{0}+2\sum_{n=1}^{\infty}b_{n}\cos\left(n(\pi-\phi)\right) \right]\mathbf{u}_z,\nonumber\\\label{E_scatt_Mie}
\end{eqnarray}
where $b_{n}$ are the well-known Mie coefficients and $\phi$ is the
angle defined between the scattering direction and the $x$-axis.

Substituting Eqs. (\ref{p}) and (\ref{m}) into Eq. (\ref{E_scatt_green}), 
the scattered field can be expressed in terms of the electric and
magnetic polarizabilities, i.e., 
\begin{eqnarray}
\textbf{E}_{s}&=&E_{0}\frac{k^{2}}{4}\left[i\alpha_{e}^{s}H_{0}\left(kr\right)+\alpha_{m}^{s}H_{1}\left(kr\right) \cos\phi\right]\hat{\mathbf{u}}_{z}\nonumber\\
&=&E_{0}\sqrt{\frac{2}{\pi kr}}e^{-i\pi/4}e^{ikr}\frac{ik^{2}}{4}\left[\alpha_{e}^{s}-\alpha_{m}^{s}\cos\phi\right]\hat{\mathbf{u}}_{z}\nonumber\\\label{E_scatt_alphas}
\end{eqnarray}
$H_{n}$ are now  the Hankel functions of the first kind, that in far-field
$kr\gg1$ read:  $H_{n}\left(kr\right)\sim\sqrt{\frac{2}{\pi kr}}e^{ikr}\left(-i\right)^{n}e^{-i\pi/4}$.

Comparing Eqs. (\ref{E_scatt_alphas}) and (\ref{E_scatt_Mie}), we
obtain, for $s$-polarization, the electric and magnetic polarizabilities in terms of the
two first Mie coefficient $b_{0}$ and $b_{1}$:
\begin{equation}
\alpha_{e}^{s}=i\frac{4}{k^{2}}b_{0},\;\;\;\;\alpha_{m}^{s}=i\frac{8}{k^{2}}b_{1}\label{alpha_e_Mie_S}
\end{equation}
It is worth remarking that these expressions characterize a dipolar cylinder in the wide sense, i.e. beyond the aforementioned Rayleigh limit of small particle as they are valid whenever the scattering
cross section of the cylinder is fully characterized by $b_{0}$ and $b_{1}$.
Notice also that the magnetic polarizability $\alpha_{m}^{s}$ of (30) corrects by a factor $2$  that given  in reference \cite{vynck2009alldielectric}.

We emphasize that these electric and magnetic polarizabilities are
 expressed in the form of Eqs. (26)-(27) with their static values being:
\begin{equation}
\alpha_{e,0}^{s}=\frac{4}{k^{2}}\frac{J_{0}\left(mx\right)J'_{0}\left(x\right)-mJ'_{0}\left(mx\right)J_{0}\left(x\right)}{J_{0}\left(mx\right)Y'_{0}\left(x\right)-mJ_{0}'\left(mx\right)Y_{0}\left(x\right)},
\end{equation}
\begin{equation}
\alpha_{m,0}^{s}=\frac{8}{k^{2}}\frac{J_{1}\left(mx\right)J'_{1}\left(x\right)-mJ'_{1}\left(mx\right)J_{1}\left(x\right)}{J_{1}\left(mx\right)Y'_{1}\left(x\right)-mJ_{1}'\left(mx\right)Y_{1}\left(x\right)}
\end{equation}
where $m=n_p$,  and $J_{n}$ $\left(x\right)$ and $Y_{n}$ $\left(x\right)$ are
the Bessel functions of the first and second kind respectively.

\subsection{P-polarization}

Let us now consider the magnetic field polarized along the cylinder
axis, i.e., $\mathbf{H}_{in}=H_{0}e^{ikx}\hat{\mathbf{u}}_{z}$.
For this polarization, the induced electric and magnetic dipoles  can
be expressed
\begin{eqnarray}
\mathbf{p}\left(0,\omega\right) & = & \varepsilon_{0}\alpha_{e}^{p}E_{0}\hat{\mathbf{u}}_{y},\label{p-1}\\
\mathbf{m}\left(0,\omega\right) & = & \alpha_{m}^{p}H_{0}\hat{\mathbf{u}}_{z},\label{m-1}
\end{eqnarray}
where $\alpha_{e,m}^{p}$ are the expressions for the dynamic polarizabilities \cite{LAKHTAKIA, madrazo, antonio, ChaumetPRB2000}
\begin{equation}
\alpha_{e}^{p}=\frac{\alpha_{e,0}^{p}}{1-\frac{ik^{2}\alpha_{e,0}^{p}}{8}},
\end{equation}
\begin{equation}
\alpha_{m}^{p}=\frac{\alpha_{m,0}^{p}}{1-\frac{i\alpha_{m,0}^{p}k^{2}}{4}},
\end{equation}

In the Rayleigh limit $\alpha_{e,0}^{p}=2\pi r_0^{2}\frac{\varepsilon_{p}-1}{\varepsilon_{p}+1},\label{alpha_e_dynamic_P}$ $\alpha_{m,0}^{p}=\pi r_0^{2}\left(\mu_{p}-1\right).\label{alpha_m_dynamic_P}$

On the other hand, analogously to the previous subsection, the scattered
field in terms of the Mie coefficients $a_{n}$ is 
\begin{eqnarray}
&&\mathbf{H}_{s}\left(\mathbf{\boldsymbol{\rho}},\boldsymbol{\rho}_{0},\omega\right)=\nonumber\\
&&-H_{0}\sqrt{\frac{2}{\pi kr}}e^{-i\pi/4}e^{ikr}\left[a_{0}+2\sum_{n=1}^{\infty}a_{n}\cos\left(n(\pi-\phi)\right) \right]\hat{\mathbf{u}}_{z},\nonumber\\\label{H_scatt_Mie}
\end{eqnarray}
Using the Green function we obtain  for the
scattered field in terms of the dynamic polarizabilities: 
\begin{eqnarray}
\mathbf{H}_{s} & = & H_{0}\frac{k^{2}}{4}\left[i\alpha_{m}^{p}H_{0}\left(kr\right)+\alpha_{e}^{p}H_{1} \left(kr\right)\cos\phi\right]\hat{\mathbf{u}}_{z}\nonumber\\
 & = & H_{0}\sqrt{\frac{2}{\pi kr}}e^{-i\pi/4}e^{ikr}\frac{ik^{2}}{4}\left[\alpha_{m}^{p}-\alpha_{e}^{s}\cos\phi\right]\hat{\mathbf{u}}_{z}.\nonumber\\\label{H_scatt_alphas}
\end{eqnarray}
From this latter expression and Eq. (\ref{H_scatt_Mie})
we write, for
$p-$polarization,the electric and magnetic polarizabilities
in terms of the two first Mie cofficients $a_{0}$ and $a_{1}$ 
\begin{equation}
\alpha_{e}^{p}=i\frac{8}{k^{2}}a_{1},\;\;\;\;\alpha_{m}^{p}=i\frac{4}{k^{2}}a_{0}\label{alpha_e_Mie_P}
\end{equation}
To obtain the expression of the polarizabilities in Gaussian units, sometimes used in the literature, one must divide Eq. (\ref{alpha_e_Mie_P}) by $4\pi$, (see also Eq. (\ref{alpha_cylinder}) of the main text). In addition, one can also write the polarizabilities
using Eqs. (35)-(36) with the static values


\begin{equation}
\alpha_{e,0}^{p}=\frac{8}{k^{2}}\frac{mJ'_{1}\left(x\right)J_{1}\left(mx\right)-J_{1}\left(x\right)J'_{1}\left(mx\right)}{mJ_{1}\left(mx\right)Y'_{1}\left(x\right)-J_{1}'\left(mx\right)Y_{1}\left(x\right)}
\end{equation}

\begin{equation}
\alpha_{m,0}^{p}=\frac{4}{k^{2}}\frac{mJ'_{0}\left(x\right)J_{0}\left(mx\right)-J_{0}\left(x\right)J'_{0}\left(mx\right)}{mJ_{0}\left(mx\right)Y'_{0}\left(x\right)-J_{0}'\left(mx\right)Y_{0}\left(x\right)}
\end{equation}

\section*{Acknowledgments}
Work supported by the Spanish  Ministerio de Economia y Competitividad (MINECO) through FIS2012-36113-C03-03  research grant.   JMA   acknowledges a  MINECO   fellowship. We  thank an anonimous reviewer for valuable comments.


\end{document}